\documentclass[onecolumn,showpacs,preprintnumbers,amsmath,amssymb,5pt]{revtex4}

\usepackage{graphicx}
\usepackage{dcolumn}
\usepackage{bm}
\usepackage{multirow}
\usepackage{indentfirst}

\begin{document}

\title{Hamiltonian Dynamics of Saturated Elongation in Amyloid Fiber Formation}

\author{Liu Hong$^{1,2}$, Xizhou Liu$^2$, Thomas C. T. Michaels$^2$, Tuomas P. J. Knowles$^2$}
\affiliation{$^1$School of Mathematics, Sun Yat-sen University, Guangzhou, 510275, P.R.C.\\
$^2$Department of Chemistry, University of Cambridge, Lensfield Road, Cambridge CB2 1EW, U.K.}

\begin{abstract}
Elongation is a fundament process in amyloid fiber growth, which is normally characterized by a linear relationship between the fiber elongation rate and the monomer concentration. However, in high concentration regions, a sub-linear dependence was often observed, which could be explained by a universal saturation mechanism. In this paper, we modeled the saturated elongation process through a Michaelis-Menten like mechanism, which is constituted by two sub-steps -- unspecific association and dissociation of a monomer with the fibril end, and subsequent conformational change of the associated monomer to fit itself to the fibrillar structure. Typical saturation concentrations were found to be $7-70\mu M$ for A$\beta$40, $\alpha$-synuclein and \textit{etc.}. Furthermore, by using a novel Hamiltonian formulation, analytical solutions valid for both weak and strong saturated conditions were constructed and applied to the fibrillation kinetics of $\alpha$-synuclein and silk fibroin.
\end{abstract}

\maketitle

\section{Introduction}

Elongation, a process of incorporating free protein molecules (or monomers) into fibrillar aggregates
through sequential monomer association with fibril ends, is considered as
the most fundamental step in amyloid fiber formation. As early as the pioneer works by Oosawa and his colleagues on actin formation in the
late 1950s \cite{oosawa1959g}, the elongation process had been identified. They found that initial
growth rate of actins varied linearly with the monomer concentration, implying monomeric subunits were added to actin filaments and made them to grow.
The fact that fiber elongation has a first-order concentration dependence on both monomeric and fibril species has been verified by plenty of following studies \cite{oosawa1975thermodynamics, collins2004mechanism},
revealing a universal bimolecular mechanism of fiber growth.

However, in experiments, a sub-linear dependence of the fiber elongation rate on the monomer concentration was often observed \cite{collins2004mechanism,lomakin1996nucleation,buell2014solution,lorenzen2012role},
especially in the regime of high concentrations. In fact, this is a rather universal phenomenon and has a deep physical basis on
saturation. It is imaginable, in the presence of too many monomers competing for the same binding site at the same time,
the fiber end will appear to be ``saturated'' since the incorporation of each monomer requires certain amount of time
and can not be finished at once. As a consequence, the elongation process appears to be blind to the instantaneous
monomer concentration in the system and shows a sub-linear dependence.

Mathematically, the process of saturated elongation could be modeled through a Michaelis-Menten like mechanism for enzyme kinetics\cite{michaelis2007kinetik}, which
includes two sub-steps -- unspecific association and dissociation of a monomer with the fibril end, and subsequent
conformational change of the associated monomer to fit itself into the fibrillar structure. Compared to the monomer association step, which is diffusion limited, the conformation change in the second process is usually
much slower and rate-limiting. In principle, all monomer-dependent processes could be saturated once the monomer concentration exceeds certain
threshold. And large amyloid proteins are more prone to get saturated than smaller ones under the same condition,
since the former generally requires a longer time to fit itself to the fibrillar structure.

The complexity of the elongation process has been extensively explored in the literature. By monitoring the deposition
of soluble A$\beta$ onto amyloid in AD brain tissue or synthetic amyloid fibrils, Esler \textit{et al.} \cite{esler2000alzheimer} showed that the A$\beta$ elongation was mediated
by two distinct kinetic processes. In the first ``dock'' phase, A$\beta$ addition to the amyloid template was fully
reversible; while in the second ``lock'' phase, the deposited peptide became irreversibly associated with the template in a time-dependent manner. A similar conclusion was reached by Scheibel \textit{et al.} \cite{scheibel2004elongation}. They examined how nuclei mediated the conversion of soluble NM domain
of Sup35 to the amyloid form. By creating single-cysteine substitution
mutants at different positions of NM domain to provide unique binding sites for various probes, the fiber elongation
was identified as a two-step process involving the capture of an intermediate, followed by its conformational
conversion. The ``dock-lock'' mechanism for amyloid fiber elongation was also explored through MD simulations. Nguyen \textit{et al.} \cite{nguyen2007monomer} simulated the formation of $(\textrm{A}\beta_{16-22})_n$ by adding a monomer to a preformed $(\textrm{A}\beta_{16-22})_{n-1}$ ($n=4-6$) oligomer. In their case, they found a rapid ``dock'' phase ($\approx50ns$) and a much slower ``lock'' phase (longer than $300ns$).

Although the physical origin of saturation during fiber elongation became clear nowadays, there were few results on the aspects of mathematical modeling and analysis, due to the intrinsic difficulty in the presence of saturation. Motivated by recent Hamiltonian formation for amyloid fiber formation by Michaels \textit{et al.} \cite{Michaels2016Hamiltonian}, we seeked to extend the Hamiltonian formation to include the saturated elongation and thus derived approximate solutions which were applicable to both saturated and non-saturated cases. Based on our analytical solutions, saturation was found to be a universal phenomenon under some conditions for a wide range of fibrous systems, including A$\beta$40, NM domain of Sup35, S6 mutants, $\alpha$-synuclein, silk fibroin proteins and \textit{etc.}.

\section{Results}
\subsection{Hamiltonian formulation of saturated elongation}
Without loss of generality, we start with a model including primary nucleation, secondary nucleation and saturated elongation \cite{knowles2009analytical,hong20174}, in which the monomer concentration $m$ and the number concentration of fibrils $P$ evolve according to
\begin{eqnarray}
&&\frac{d}{dt}m=-2k_+\frac{m}{1+m/K_e}P,\label{kineticmodel}\\
&&\frac{d}{dt}P=k_nm^{n_c}+k_2m^{n_2}(m_{tot}-m),\label{kineticmodel2}
\end{eqnarray}
where $m_{tot}$ is the total protein concentration. $k_n$, $k_2$  and $k_+$  denote the reaction rates for homogeneous primary nucleation, secondary nucleation and saturated elongation. $n_c$ and $n_2$ represent the critical nucleus size for primary nucleation and secondary nucleation respectively. Note, in above equation, we have surface catalyzed secondary nucleation for $n_2>0$ and fragmentation dominant secondary nucleation for $n_2=0$.

The term $m/(1+m/K_e)$ for saturated elongation appeared on the right hand side of Eq. \ref{kineticmodel} can be derived from a more comprehensive consideration about the elongation process (see Methods for details). The Michaelis constant $K_e$ acts as an index of the effective monomer concentration in the saturated elongation model. If the monomer concentration is much higher than the Michaelis constant $m\gg K_e$, we have $m/(1+m/K_e)\approx K_e$ meaning only a constant concentration $K_e$ could be used for fiber elongation, a key feature of the saturation phenomenon; contrarily, if the monomer concentration is far lower than the Michaelis constant $m\ll K_e$, we have $m/(1+m/K_e)\approx m$ and recover the classical elongation process as expected. Especially, when the monomer concentration is equal to the Michaelis constant $m=K_e$, the rate of elongation is half of its maximal value.

To reformulate the saturated elongation model into a Hamiltonian structure \cite{Michaels2016Hamiltonian}, we introduce generalized coordinates for momentum and position as $p=2k_+P$  and $q=\ln(m_{tot}/m)+(m_{tot}-m)/K_e$. Then the monomer concentration can be expressed as $m=K_e\cdot W\big[(m_{tot}/K_e)\exp(m_{tot}/K_e)\exp(-q)\big]$, in which $x=W(y)$  stands for the Lambert W function solving the equation  $y=x\cdot\exp(x)$. Now Eq. 1 can be casted into the Hamiltonian structure in classical mechanics \cite{Michaels2016Hamiltonian},
\begin{eqnarray}
&&\dot{q}=\frac{\partial H}{\partial p},\\
&&\dot{p}=-\frac{\partial H}{\partial q},
\end{eqnarray}
where the Hamiltonian $H=p^2/2+V(q)$ with the potential energy
\begin{eqnarray}\label{potentialen}
V(x(q))=\mu^2\bigg[\frac{W^{n_c+1}(x)}{n_c+1}+\frac{W^{n_c}(x)}{n_c}\bigg]+\alpha\eta^2\bigg[\frac{W^{n_2+1}(x)}{n_2+1}+\frac{W^{n_2}(x)}{n_2}\bigg]-\eta^2\bigg[\frac{W^{n_2+2}(x)}{n_2+2}+\frac{W^{n_2+1}(x)}{n_2+1}\bigg],
\end{eqnarray}
in which $x(q)=(m_{tot}/K_e)\exp(m_{tot}/K_e)\exp(-q)$, $\mu=\sqrt{2k_+k_nK_e^{n_c}}$, $\eta=\sqrt{2k_+k_2K_e^{n_2+1}}$  and  $\alpha=m_{tot}/K_e$.

Further introducing the Lagrangian $L=p^2/2-V(q)$ and applying the principle of least action, we get the Euler-Lagrange equation  $\ddot{q}=-\partial_qV$. Integrating the Euler-Lagrange equation once provides an implicit solution for $q(t)$  in terms of a single integral
\begin{eqnarray}\label{ELeq}
t=\int_0^q\frac{dq'}{\sqrt{2[V(0)-V(q')]}}.
\end{eqnarray}
Above equation could be solved approximately under two limiting conditions: strong saturation $\alpha\gg1$ and weak saturation $\alpha\ll1$ (see Methods for details). To combine these two cases together, here we propose a unified solution obeying the generalized logistic form,
\begin{eqnarray}\label{massconcentration}
m/m_{tot}=\bigg[1+\frac{1}{\theta}y(1+y)^{\alpha/(1+\alpha)}\bigg]^{-\theta},
\end{eqnarray}
where $y=\epsilon\exp\big(\frac{\kappa t}{\sqrt{1+\alpha}}\big)$, $\kappa=\sqrt{2k_+k_2m_{tot}^{n_2+1}}$ and $\theta=\sqrt{2/[n_2(n_2+1)]}$. During the derivation, a critical concentration of filaments  $\epsilon\equiv M(0)/m_{tot}=k_nm_{tot}^{n_c-n_2-1}/(2k_2)\ll1$ has been introduced in the integration to seed the system, so that the resulting expression for $m(t)$  matches the leading order term of solutions for the early time. In the special case of $n_2=0$ ($\theta\rightarrow\infty$), which corresponds to fragmentation, the solution reduces to $m/m_{tot}=\exp\bigg[-y(1+y)^{\alpha/(1+\alpha)}\bigg]$ by noticing the fact $\lim_{b\rightarrow\infty}(1+a/b)^b=e^a$.

Based on above solution, we can roughly determine that the half time $t_{1/2}\approx\ln(1/\epsilon)\sqrt{1+\alpha}/\kappa$, while the apparent fiber growth rate $k_{app}\approx\kappa/\sqrt{1+\alpha}$. It is easily seen that $t_{1/2}\propto m_{tot}^{-(n_2+1)/2}$ as $\alpha\ll1$ and $t_{1/2}\propto m_{tot}^{-n_2/2}$ as $\alpha\gg1$. Especially, when $n_2=0$ and $\alpha\gg1$, we have $t_{1/2}\propto -ln(m_{tot})$.

Furthermore, by using the fact of energy conservation in Hamiltonian systems $V(0)=H(0)=H(t)=p^2(t)/2+V(t)$, a simple relation for the number concentration of fibrils could be constructed as
\begin{eqnarray}
P(t)=\frac{\sqrt{\theta^2+\alpha\vartheta^2}\kappa}{2k_+}\bigg[1-\bigg(\frac{m}{m_{tot}}\bigg)^{1/\theta}\bigg],
\end{eqnarray}
where $\vartheta=\sqrt{2/[(n_2+1)(n_2+2)]}$. Again, we need to pay attention to the case $n_2=0$, which gives $P(t)=\frac{k_n}{2}m_{tot}^{n_c}t+k_2m_{tot}\frac{\sqrt{\alpha+1}}{\kappa}\ln(\frac{1+y}{1+\epsilon})$.

\subsection{Applications to Amyloid Systems}
In principle, the elongation process for all amyloid proteins will get saturated when the monomer concentration becomes high enough. Therefore, it would be interesting to know the typical saturation concentration (or the Michaelis constant) for each protein. Here we collected data for four typical amyloid systems in the literature, \textit{i.e.} A$\beta$40, NM domain from yeast prion protein Sup35, S6 protein mutant IA8 and $\alpha$-synuclein. All of them showed clear signals of saturation within the examined concentration region (see Fig. 1).

To be specific, we looked at the relationship between the fiber elongation rate $V$ and the initial monomer concentration $m$, which could be modeled by the Michaelis-Menten equation
\begin{eqnarray}\label{MMeqn}
V=V_{max}\frac{m}{K_e+m},
\end{eqnarray}
in consistent with Eq. \ref{kineticmodel}. Two unknown parameters -- the maximal fiber elongation rate $V_{max}$ and saturation concentration $K_e$ could be directly extracted from the Lineweaver-Burk plot of the data ($1/V$ \textit{v.s.} $1/m$) \cite{lee1971enzymic}. The double reciprocal plot yields a straight line, in which the x intercept gives $-1/(V_{max}K_e)$ and the y intercept gives $1/V_{max}$. In this way, the saturation concentration $K_e$ for four amyloid proteins were determined to be from $7\mu M$ to $70\mu M$ (see Fig. S1), a relatively narrow region considering the dramatic chemical differences among these proteins.

\begin{figure}[h]
\centering
\includegraphics[width=0.8\textwidth,height=0.5\textwidth]{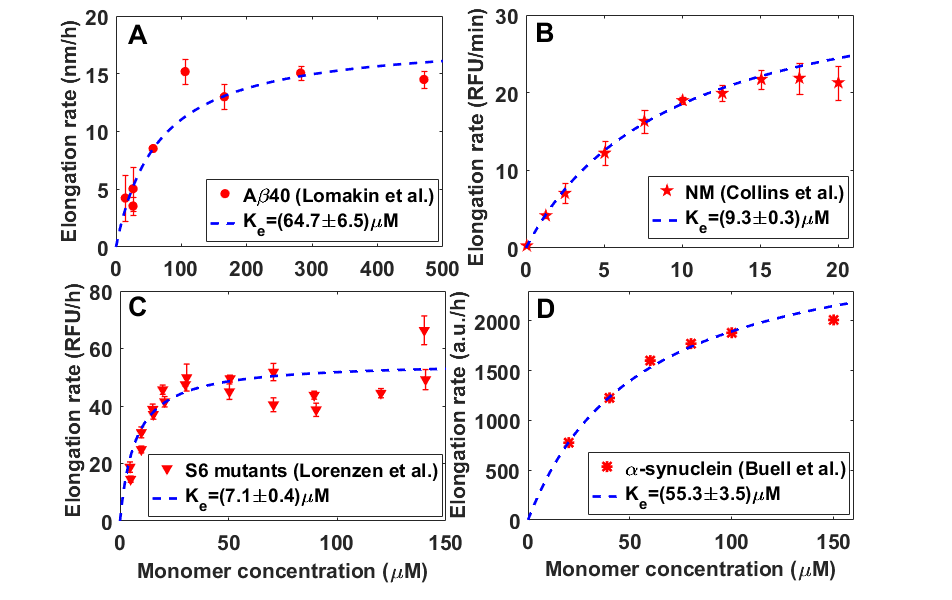}
\caption{The fiber elongation rate (or initial growth rate) shows a sub-linear dependence on monomer concentration. Data for (A) A$\beta$40 (taken from Ref. \cite{lomakin1996nucleation}) was determined through the temporal evolution of the hydrodynamic radius by QLS. Data for (B) NM domain from yeast prion protein Sup35 (Ref. \cite{collins2004mechanism}), (C) S6 mutant IA8 (Ref. \cite{lorenzen2012role}) and (D) $\alpha$-synuclein (Ref. \cite{buell2014solution}) were all measured by continuous thioflavin T assay. In each plot, the fiber elongation rate was fitted by the Michaelis-Menten equation (\ref{MMeqn}) (blue dashed lines) with the saturation concentration $K_e$ marked accordingly.}
\end{figure}

With the saturation concentration in hand, we made a further validation by examining the fibrillation kinetics of $\alpha$-synuclein under seeded condition according to  \cite{buell2014solution}. The global fitting showed fragmentation was the dominant secondary nucleation mechanism for $\alpha$-synuclein aggregation in this case (see Fig. 2). And the elongation process became saturated when the monomer concentration got close to $55.3\mu M$. Furthermore, the average length of $\alpha$-synuclein seeds was determined to be around $175$ monomers. This value was in a perfect agreement with the AFM statistics ($\sim83nm$) \cite{buell2014solution}, given single monomer size contribution to the fibril axis as $0.47nm$ \cite{van2008concentration}.

\begin{figure}[h]
\centering
\includegraphics[width=0.7\textwidth,height=0.5\textwidth]{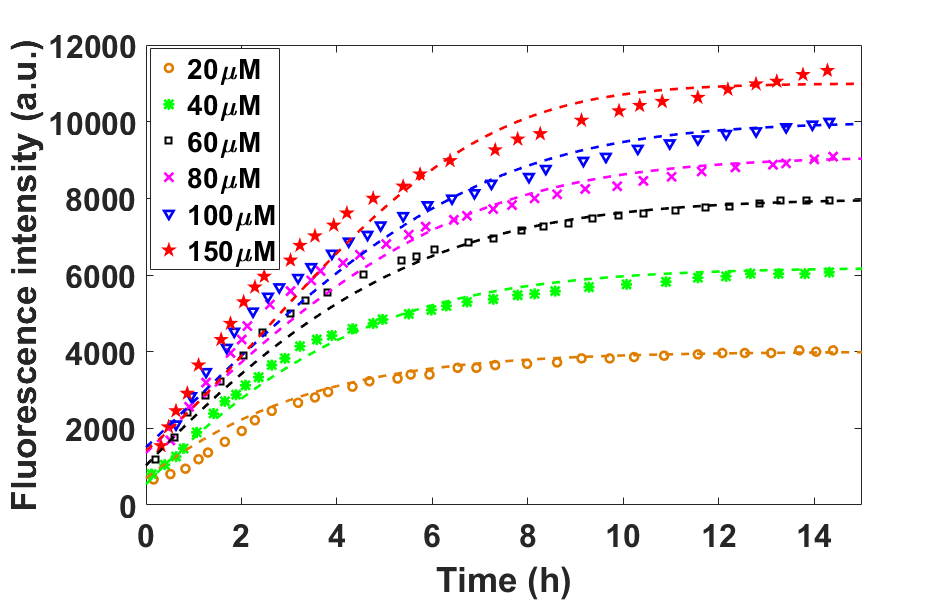}
\caption{(A) Fibrillation of $\alpha$-synuclein followed the saturated elongation mechanism (data taken from \cite{buell2014solution}). Global fittings (dashed lines) were performed under six different protein concentrations 20(brown circles), 40(green stars), 60(black squares), 80(purple crosses), 100(blue triangles), and 150$\mu M$(red pentagram) with $3.5\mu M$ sonicated seeds. Parameters were given as $k_n=2.3\times10^{-9}\mu M^{-1}h^{-1}$, $k_+=9.5\mu M^{-1}h^{-1}$, $k_2=6.0\times10^{-17}h^{-1}$, $K_e=55.3\mu M$, $n_c=3$ and $n_2=0$. The initial number concentration of seeds was determined as $P(0)=0.02\mu M$.}
\end{figure}

Silk fibroin (SF) fibrils represented another example of natural fibrous assemblies. SF fibrils shared many similarities to amyloid fibrils, however, unlike amyloid fibrils, both $\beta$-strands and $\beta$-sheets in SF fibrils were parallel to the fibril axis. To our knowledge, this structural arrangement perhaps resulted in the remarkable elasticity in this class of materials. Recently, growing attentions have been focused on utilising SF fibrils as proteinaceous building blocks in material engineering. However, understanding of the formation of such fibrous systems remained challenging, and several models have been proposed to explain the mechanistic picture of the assembly process. Herein, we proposed to use the saturated elongation model to study the aggregation process of regenerated silk fibroin, and more specifically the kinetics of the SF fibrillation under acidic environments.

For silk fibroin, the half-time of fibrillation showed a very weak scaling dependence on the monomer concentration ($t_{1/2}\propto m_{tot}^{\gamma}$ with $\gamma=-0.03\sim-0.14$), which is a key feature of saturated elongation (see Fig. 3). In contrast, it is well-known that the classical Oosawa model \cite{oosawa1975thermodynamics}, constituted by primary nucleation and elongation, predicts $\gamma=-n_c/2$ with $n_c\geq1$ as the critical nucleus size. We further have $\gamma=-1/2$ for the fragmentation dominant mechanism \cite{knowles2009analytical} and $\gamma=-(n_2+1)/2$ for the surface catalyzed secondary nucleation \cite{Ruschak2007Fiber}, where $n_2\geq0$ stands for critical nucleus size for secondary nucleation. As $\gamma<=-1/2$ in all these models, it is clear that none of them is applicable to silk fibroin.

According to the results for strong saturated elongation with fragmentation (Eq. \ref{massconcentration} with $\alpha\gg1$ and $n_2=0$), only three parameters were needed for analyzing the data, that are the critical nucleus size $n_c$, the fiber growth rate $k_+k_2K_e$ and the ratio between primary nucleation and secondary nucleation $k_n/k_2$. It turned out that, with the increase of dissolvation time, both the fiber growth rate and the ratio between primary nucleation and secondary nucleation decreased dramatically for about an order of magnitude, leading to a much smaller primary nucleation rate and fiber elongation rate given the fragmentation rate unchanged. This explained the observed apparent slowing down in SF fibrillation kinetics. Furthermore, a consistent increase in the critical nucleus size was observed. To be exact, we found $n_c\approx1.0$ for 15 minutes of dissolvation; $n_c\approx1.1$ for 30 minutes and $n_c\approx2.8$ for 60 minutes. This observation agreed with the fact that the size distribution of silk proteins shifts to low molecular weight during dissolvation and more units (monomers) are required for forming a stable nucleus with a relatively fixed surface-volume ratio.

\begin{figure}[h]
\centering
\includegraphics[width=0.7\textwidth,height=0.5\textwidth]{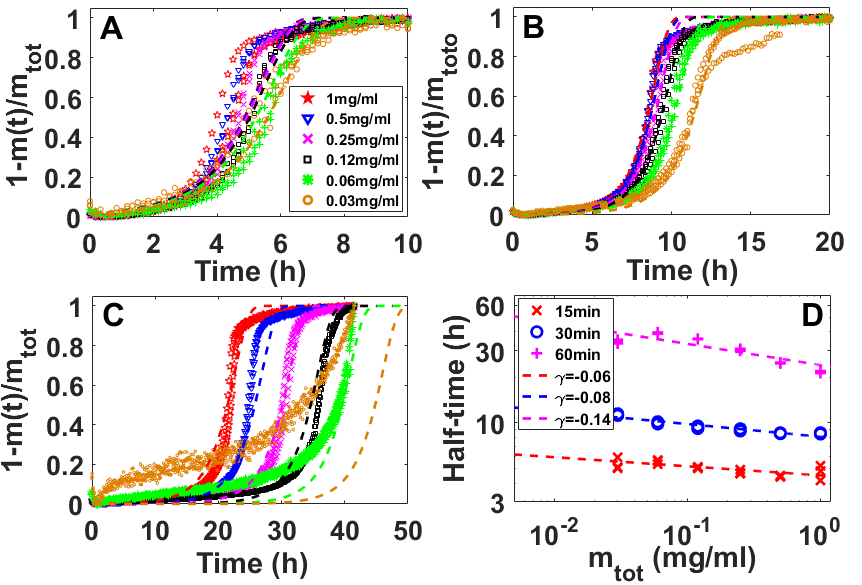}
\caption{Fibrillation of silk proteins after (A) 15, (B) 30 and (C) 60 minutes of dissolvation at pH3 (unpublished data by Xizhou Liu). Under each of six different concentrations 1(red pentagram), 0.5(blue triangles), 0.25(purple crosses), 0.12(black squares), 0.06(green stars), and 0.03$mg/ml$(brown circles), three independent measurements were repeated and showed by circles. Global fittings (dashed lines) were performed with fragmentation and strong saturation dominated mechanism ($n_2=0$ and $\alpha\gg1$). Parameters were given as $k_+k_2K_e=0.27h^{-2}$, $k_n/k_2=2.8\times10^{-2}(mg/ml)^{-n_c+1}$, $n_c=1.0$ for (A); $k_+k_2K_e=0.25h^{-2}$, $k_n/k_2=2.5\times10^{-3}(mg/ml)^{-n_c+1}$, $n_c=1.1$ for {B}; and $k_+k_2K_e=0.044h^{-2}$, $k_n/k_2=1.8\times10^{-3}(mg/ml)^{-n_c+1}$, $n_c=2.8$ for {C} respectively. (D) Half-time for silk fibroin fibrillation v.s. monomer concentration. The best fitted scaling exponents ($t_{1/2}\propto m_{tot}^{\gamma}$) were marked on the plot.}
\end{figure}

\section{Methods}

\subsection{Derivation of saturated elongation model}
The process of elongation could be modeled through a Michaelis-Menten like mechanism for enzyme kinetics \cite{michaelis2007kinetik}, which includes two sub-steps -- unspecific association and dissociation of monomers $A$ with the fibril ends, and conformational change of monomers after association,
\begin{eqnarray}
P_{free}+A\mathop{\rightleftharpoons}^{k_a^+}_{k_a^-}P_{bound}\mathop{\rightarrow}^{k_c}P_{free},
\end{eqnarray}
which can be modeled through following equations,
\begin{eqnarray}
&&\frac{d}{dt}m=-2k_cP_{bound},\\
&&\frac{d}{dt}P_{bound}=2k_a^+mP_{free}-2k_a^-P_{bound}-2k_cP_{bound},
\end{eqnarray}
where $P_{free}+P_{bound}=P$.

Above model can be further simplified based on the Quasi Stead-State Approximation (QSSA). QSSA assumes the monomer-attached fibril ends $P_{bound}$ are always in a dynamical balance, which means that we can take terms on the right-hand side of the second equation to be zero, \textit{i.e.}
\begin{eqnarray}
0=2k_a^+mP_{free}-2k_a^-P_{bound}-2k_cP_{bound},
\end{eqnarray}
which give formulas like the Michaelis-Menten reactions,
\begin{eqnarray}
&&P_{free}=P\bigg(1+\frac{m}{K_e}\bigg)^{-1},\\
&&P_{bound}=P\frac{m}{K_e}\bigg(1+\frac{m}{K_e}\bigg)^{-1},
\end{eqnarray}
where the Michaelis constant is given by $K_e=(k_a^-+k_c)/k_a^+$. Consequently, the model for saturated elongation becomes
\begin{eqnarray}
\frac{d}{dt}m=-2k_+Pm\bigg(1+\frac{m}{K_e}\bigg)^{-1},
\end{eqnarray}
where $k_+=k_c/K_e=k_c k_a^+/(k_a^-+k_c)$.

\subsection{Non-physical solutions for full saturation model}
In the limit of strong saturation $\alpha\gg1$, the model for saturated elongation can be simplified as
\begin{eqnarray}
&&\frac{d}{dt}m=-2k_+K_eP,\\
&&\frac{d}{dt}P=k_nm^{n_c}+k_2m^{n_2}(m_{tot}-m).
\end{eqnarray}
However, above equations have non-physical solutions and can not be considered as a proper model in the long time. This is what we are going to address here.

Take time derivatives on both sides of the first equation and use the second one ro replace the term $dP/dt$, we get
\begin{eqnarray}
&&\frac{d^2}{dt^2}m=-2k_+K_e\big[k_nm^{n_c}+k_2m^{n_2}(m_{tot}-m)\big].
\end{eqnarray}
Specifically, by letting $k_2=0$ and $n_c=1$, above equation can be mapped to 1-d harmonic oscillator model with the solution $m/m_{tot}=\cos(\sqrt{2k_+k_nK_e}t)$. Actually, it is not difficult to show that the general solution is oscillatory when $n_c$ and $n_2$ are odd and monotonically decreasing to negative infinity when both are even. Those solution are non-physical and valid only within a finite time region before $m(t)=0$. Contrarily, the original model without replacing by $K_e$ does not suffer from this kind of drawbacks, meaning the non-physical solutions are mainly caused by the improper oversimplification of the model by the condition of full saturation.

\subsection{Approximate solutions for the Euler-Lagrange equation}
In this section, we are going to present the details on solving the Euler-Lagrange equation in Eq. \ref{ELeq}. Before starting, we first look at the potential energy given in Eq. \ref{potentialen}. Since in general the secondary nucleation is stronger than primary nucleation $\alpha\eta^2\gg\mu^2$, the first term in the the potential energy can be neglected, which results in
\begin{eqnarray}\label{energydifference}
&&V(0)-V(q')\approx\alpha\eta^2\bigg(\frac{\alpha^{n_2+1}}{n_2+1}+\frac{\alpha^{n_2}}{n_2}\bigg)-\eta^2\bigg(\frac{\alpha^{n_2+2}}{n_2+2}+\frac{\alpha^{n_2+1}}{n_2+1}\bigg)-\alpha\eta^2\bigg(\frac{W^{n_2+1}}{n_2+1}+\frac{W^{n_2}}{n_2}\bigg)+\eta^2\bigg(\frac{W^{n_2+2}}{n_2+2}+\frac{W^{n_2+1}}{n_2+1}\bigg)\nonumber\\
&&=\kappa^2\bigg[\frac{\tilde{W}^{n_2+1}}{n_2+1}-\frac{\tilde{W}^{n_2}}{n_2}+\frac{1}{n_2(n_2+1)}\bigg]+\alpha\kappa^2\bigg[\frac{\tilde{W}^{n_2+2}}{n_2+2}-\frac{\tilde{W}^{n_2+1}}{n_2+1}+\frac{1}{(n_2+1)(n_2+2)}\bigg]\nonumber\\
&&=\frac{\kappa^2}{n_2(n_2+1)}\big[n_2\tilde{W}^{n_2+1}-(n_2+1)\tilde{W}^{n_2}+1\big]+\frac{\alpha\kappa^2}{(n_2+1)(n_2+2)}\big[(n_2+1)\tilde{W}^{n_2+2}-(n_2+2)\tilde{W}^{n_2+1}+1\big]\nonumber\\
&&\approx\frac{\kappa^2\theta^2}{2}\big(1-\tilde{W}^{1/\theta}\big)^2+\frac{\alpha\kappa^2\vartheta^2}{2}\big(1-\tilde{W}^{1/\vartheta}\big)^2,
\end{eqnarray}
where  $\tilde{W}=W/\alpha\in[0,1]$, $\kappa=\sqrt{2k_+k_2m_{tot}^{n_2+1}}$, $\theta=\sqrt{2/[n_2(n_2+1)]}$ and $\vartheta=\sqrt{2/[(n_2+1)(n_2+2)]}$. The first approximation omits the contribution of primary nucleation; the second one is referred to the argument by Michaels \textit{et al.} \cite{Michaels2016Hamiltonian}.

Now Eq. \ref{ELeq} could can be solved under two limiting conditions:

\textbf{(1)	Strong saturation $\alpha\gg1$.} In this case, the first term in the potential energy difference can be neglected comparing to the second one, thus
\begin{eqnarray}
t&=&\int_0^q\frac{dq'}{\sqrt{2[V(0)-V(q')]}}\approx\int\frac{dq'}{\sqrt{\alpha\kappa^2\vartheta^2\big(1-\tilde{W}^{1/\vartheta}\big)^2}}=\frac{-1}{\sqrt{\alpha}\kappa\vartheta}\int\frac{dx}{x[1-\tilde{W}^{1/\vartheta}(x)]}\nonumber\\
&=&\frac{-1}{\sqrt{\alpha}\kappa\vartheta}\int_1^{m/m_{tot}}\frac{(1+\alpha\tilde{W})d\tilde{W}}{\tilde{W}(1-\tilde{W}^{1/\vartheta})}=\frac{-1}{\sqrt{\alpha}\kappa\vartheta}\int\frac{[1+\alpha(1+\tilde{f})^{-\vartheta}]d[(1+\tilde{f})^{-\vartheta}]}{(1+\tilde{f})^{-\vartheta}[1-(1+\tilde{f})^{-1}]}\nonumber\\
&=&\frac{1}{\sqrt{\alpha}\kappa}\bigg[\int\frac{d\tilde{f}}{\tilde{f}}+\alpha\int\frac{df}{\tilde{f}(1+\tilde{f})^{\vartheta}}\bigg]\approx\frac{1}{\sqrt{\alpha}\kappa}\bigg[\int\frac{d\tilde{f}}{\tilde{f}}+\alpha \int\frac{d\tilde{f}}{\tilde{f}(1+\vartheta \tilde{f})}\bigg]\nonumber\\
&\approx&\frac{1}{\sqrt{\alpha}\kappa}[(\alpha+1)\ln \tilde{f}-\alpha\ln(1+\vartheta \tilde{f})]\big|_{\epsilon/\vartheta}^{(m/m_{tot})^{-1/\vartheta}-1}=\frac{1}{\sqrt{\alpha}\kappa}\bigg[(\alpha+1)\ln \big(\frac{\vartheta f}{\epsilon}\big)-\alpha\ln\big(\frac{1+\vartheta f}{1+\epsilon}\big)\bigg].
\end{eqnarray}
During the derivation, we use the formula $dW=\frac{W(x)}{x[1+W(x)]}dx$ for the Lambert W function and replacements $\tilde{W}=(1+\tilde{f})^{-\vartheta}$ and $m/m_{tot}=(1+f)^{-\vartheta}$. In the second approximation, we take $(1+\tilde{f})^{\vartheta}\approx(1+\vartheta \tilde{f})$ through Taylor expansion since the main contribution to the integral comes from the part of $\tilde{f}\rightarrow0$. In the third one, a critical concentration of filaments $\epsilon\equiv M(0)/m_{tot}=k_nm_{tot}^{n_c-n_2-1}/(2k_2)\ll1$ has been introduced in the integration to seed the system and aslo avoid the singularity at $\tilde{f}=0$.

As $\alpha\gg1$, $\alpha/(\alpha+1)\approx1$. We solve the function from the formula
\begin{eqnarray}
f=\frac{\epsilon}{\vartheta}\exp\bigg(\frac{\sqrt{\alpha}\kappa t}{\alpha+1}\bigg)\cdot\bigg(\frac{1+\vartheta f}{1+\epsilon}\bigg)^{\frac{\alpha}{\alpha+1}}\approx\frac{\epsilon}{\vartheta}\exp\bigg(\frac{\sqrt{\alpha}\kappa t}{\alpha+1}\bigg)\cdot\bigg(\frac{1+\vartheta f}{1+\epsilon}\bigg),
\end{eqnarray}
which gives
\begin{eqnarray}
f=\frac{1}{\vartheta}(1-z)^{-1}-\frac{1}{\vartheta},
\end{eqnarray}
where $z=\frac{\epsilon}{(1+\epsilon)}\exp(\frac{\sqrt{\alpha}\kappa t}{\alpha+1})$. It is noted that the above solution is valid only within the region $t\in[0,\ln(\frac{1+\epsilon}{\epsilon})\frac{\alpha+1}{\sqrt{\alpha}\kappa})$, in consistency with the emergency of non-physical solutions under the condition of full saturation we discussed before. To extend the solution to the whole time regime, a simple way is to go through the Taylor expansion
\begin{eqnarray}
f=\frac{1}{\vartheta}\big(z+z^2+z^3\cdots\big)\approx\frac{1}{\vartheta}\bigg(\frac{z^{n+1}-z}{z-1}\bigg).
\end{eqnarray}
Again, the Taylor series does not converge when $z\geq1$, so we should bear in mind that only finite terms could be kept during the calculation. Finally, we have
\begin{eqnarray}
\frac{m}{m_{tot}}=\bigg[1+\frac{1}{\vartheta}\bigg(\frac{z^{n+1}-z}{z-1}\bigg)\bigg]^{-\vartheta}.
\end{eqnarray}
Note, under the condition of strong saturation, there is no singular problem for the fragmentation case when $n_2=0$.

\textbf{(2)	Weak saturation $\alpha\ll1$.} In this case, the first term is kept while the second one will be omitted in Eq. \ref{energydifference}, which leads to
\begin{eqnarray}
t\approx\int_0^q\frac{dq'}{\sqrt{\kappa^2\theta^2\big(1-\tilde{W}^{1/\theta}\big)^2}}.
\end{eqnarray}
Following the same derivation as above, we get
\begin{eqnarray}
f=\frac{\epsilon}{\theta}\exp\bigg(\frac{\kappa t}{\alpha+1}\bigg)\cdot\bigg(\frac{1+\theta f}{1+\epsilon}\bigg)^{\frac{\alpha}{\alpha+1}}\approx\frac{\epsilon}{\theta}\exp\bigg(\frac{\kappa t}{\alpha+1}\bigg),
\end{eqnarray}
as $\frac{\alpha}{\alpha+1}\approx0$. Therefore
\begin{eqnarray}
m/m_{tot}=\bigg[1+\frac{\epsilon}{\theta}\exp\bigg(\frac{\kappa t}{\alpha+1}\bigg)\bigg]^{-\theta}.
\end{eqnarray}
Especially, when $n_2=0$, $\theta=\sqrt{2/[n_2(n_2+1)]}\rightarrow\infty$. The limit gives
\begin{eqnarray}
m/m_{tot}=\exp\bigg[-\frac{\epsilon}{\theta}\exp\bigg(\frac{\kappa t}{\alpha+1}\bigg)\bigg].
\end{eqnarray}

\textbf{(3) Constructing a universal solution.} Now a major task is how to combine the solutions under two different limiting codnitions and construct a universally valid solution for all $\alpha$. Considering similarities and differences of solutions for both $\alpha\gg1$ and $\alpha\ll1$, we suggest the following formula as a candidate
\begin{eqnarray}
m/m_{tot}=\bigg[1+\frac{1}{\theta}y(1+y)^{\alpha/(1+\alpha)}\bigg]^{-\theta},
\end{eqnarray}
in which $y=\epsilon\exp\big(\frac{\kappa t}{\sqrt{1+\alpha}}\big)$. A direct comparison with numerical solutions were given in Fig. 4.

Furthermore, using the fact of energy conservation in Hamiltonian systems $V(0)=H(0)=H(t)=p^2(t)/2+V(t)$, a simple relation for the number concentration of fibrils can be deduced as
\begin{eqnarray}
P(t)=\frac{\kappa}{2k_+}\sqrt{\theta^2\bigg[1-\bigg(\frac{m}{m_{tot}}\bigg)^{1/\theta}\bigg]^2+\alpha\vartheta^2\bigg[1-\bigg(\frac{m}{m_{tot}}\bigg)^{1/\vartheta}\bigg]^2}\approx\frac{\sqrt{\theta^2+\alpha\vartheta^2}\kappa}{2k_+}\bigg[1-\bigg(\frac{m}{m_{tot}}\bigg)^{1/\theta}\bigg],
\end{eqnarray}
if we neglect the difference between $\theta$ and $\vartheta$ in the scaling exponents. Notice that when $n_2\rightarrow0$, above formula will break down. We suggest to integral Eq. \ref{kineticmodel2} directly instead, which gives $P(t)=\frac{k_n}{2}m_{tot}^{n_c}t+k_2m_{tot}\frac{\sqrt{\alpha+1}}{\kappa}\ln(\frac{1+y}{1+\epsilon})$ by taking $m/m_{tot}=(1+y)^{-1}$.

\begin{figure}[h]
\centering
\includegraphics[width=0.7\textwidth,height=0.5\textwidth]{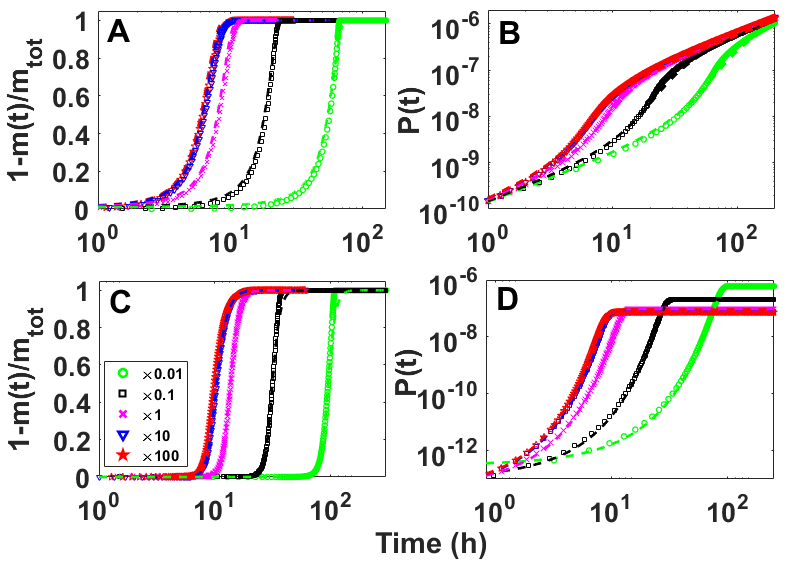}
\caption{Validation of approximate solutions in Eqs. 7 and 8 (dashed lines) by directly comparing to numerical solutions of Eqs. 1 and 2 (symbols). (A, B) Parameters $n_c=2, n_2=0, k_+=10^4M^{-1}s^{-1}, k_2=9.8\times10^{-8}s^{-1},k_n=9.8\times10^{-5}M^{-1}s^{-1}$, $m_{tot}=20\mu M$ were determined through the best global fitting of kinetic data of yeast prion Ure2p from \cite{Zhu2003Relationship}. (C, D) Parameters $n_c=8, n_2=4, k_+=10^4M^{-1}s^{-1}, k_2=6.4\times10^{6}M^{-4}s^{-1},k_n=6.1\times10^{9}M^{-7}s^{-1}$, $m_{tot}=700\mu M$ were determined through the IAPP data in \cite{Ruschak2007Fiber}. In both cases, in order to cover both weakly and strongly saturated conditions, the Michaelis constant $K_e$ was varied by five orders of magnitude in order to let $\alpha$ go from $10^{-2}$ to $10^2$.}
\end{figure}

\subsection{Experimental protocols}
\textbf{(1) Reconstitution process.}

Bombyx mori cocoons were reconstituted into fibroin solutions as previously published. In brief, small pieces of silkworm cocoons (Mindset, UK) were degummed (removal of sericin) through boiling in $0.02M$ sodium carbonate solution (typically $5g$ of cocoons to $2L$ of water). Boiling times were $15$, $30$ and $60$ minutes separately. We used the different boiling time as models for large, medium and small MW fragments of the SF as reported in \cite{partlow2016silk}. After the removal of sericin, the resulted fibers were rinsed with MQ water and air-dried. The fibers were then dissolved in $9.3M$ Lithium bromide solution at $65^oC$ for $4h$ (typically at $20 w/v\%$). The resulted fibroin solution was then dialysed against MQ water for $48h$ to remove the Lithium Bromide salt.

\textbf{(2) ThT assay.}

Stock solution of fibroin (ca. $40mg/ml$) was diluted with phosphate buffer (pH3) to obtain the final concentrations as indicated. The final concentration of ThT used was $50\mu M$. SF solutions were then incubated at $37^oC$ and the self-assembly kinetics was examined in a fluorescent plate reader Fluorostar (BMG Labtech) using a ThT filter (excitation $440nm$/emission $490nm$).

\section{Conclusion}
Elongation is a fundament process in amyloid fiber growth, which is normally characterized by a linear relationship between the fiber elongation rate and the available monomer concentration. However, in high concentration regions, a sub-linear dependence was often observed, which could be explained by a universal saturation mechanism. Typical saturation concentrations for amyloid fiber elongation were found to be $7-70\mu M$, based on analysis of four typical amyloid systems, including A$\beta$40 and $\alpha$-synuclein.

The saturated elongation process could be modeled through a Michaelis-Menten like mechanism for enzyme kinetics, which is constituted by two sub-steps -- unspecific association and dissociation of a monomer with the fibril end, and subsequent conformational change of the associated monomer to fit itself to the fibrillar structure. Through a Hamiltonian formulation, analytical solutions valid for both weak and strong saturated conditions were constructed, compared with numerical solutions and applied to the fibrillation kinetics of $\alpha$-synuclein and silk fibroin. We hope our results will draw attentions to the experimental design and data analysis in amyloid studies, especially when dealing with high protein concentrations.

\section*{Acknowledgment}
L. H. acknowledges the financial supports from the National Natural Science Foundation of China (Grant 21877070) and the Hundred-Talent Program of Sun Yat-Sen University.

\bibliography{saturation}{}
\bibliographystyle{unsrt}

\end{document}